\documentclass[runningheads]{llncs}
\usepackage{graphicx}
\usepackage{amsmath}
\usepackage{amsfonts}

\usepackage{xcolor}
\definecolor{mygreen}{RGB}{0,159,57}
\definecolor{myyellow}{RGB}{255,200,0}
\definecolor{myorange}{RGB}{255,140,0}
\definecolor{myblue}{RGB}{0,50,255}

\newcommand{\vampire}{\textsc{Vampire}}
\newcommand{\avatar}{\textsc{Avatar}}
\newcommand{\smtlib}{\textsc{smt-lib}}

\newcommand{\thax}{\mathit{thAx}}
\newcommand{\allax}{\mathit{allAx}} 
\newcommand{\ratio}{\mathit{frac}}

\usepackage{siunitx}
\usepackage{url}

\makeatletter
\newcommand\footnoteref[1]{\protected@xdef\@thefnmark{\ref{#1}}\@footnotemark}
\makeatother

\begin{document}

\title{Layered Clause Selection for Theory Reasoning\\(short paper)}



\author{Bernhard Gleiss\inst{1}
\and
Martin Suda\inst{2}
}

\authorrunning{B. Gleiss and M. Suda}

\institute{TU Wien, Austria \and
Czech Technical University in Prague, Czech Republic}
\maketitle              

\begin{abstract}
Explicit theory axioms are added by a saturation-based theorem prover as one of the techniques for supporting theory reasoning.
While simple and effective, adding theory axioms can also pollute the search space with many irrelevant consequences.
As a result, the prover often gets lost in parts of the search space where the chance to find a proof is low.
In this paper, we describe a new strategy for controlling the amount of reasoning with explicit theory axioms.
The strategy refines a recently proposed two-layer-queue clause selection 
and combines it with a heuristic measure of the amount of theory reasoning in the derivation of a clause.
We implemented the new strategy in the automatic theorem prover \vampire{} and
present an evaluation showing that our work 
dramatically improves the state-of-the-art clause-selection strategy in the presence of theory axioms.
\end{abstract}

\section{Introduction}

Thanks to recent advances, 
saturation-based theorem provers are increasingly used to reason about problems requiring quantified theory-reasoning \cite{gleiss-fmcad19,tiros-aws}.
One of the standard techniques to enable such reasoning is
to automatically add first-order axiomatisations of theories detected in the input \cite{SPASS_T,DBLP:conf/cav/KovacsV13}.
For example, (incomplete) axiomatisations of integer and real arithmetic 
or McCarthy's axioms of the theory of arrays \cite{Mccarthy62towardsa} are routinely used.
%
%
While this simple technique is often effective,
we observed (see also \cite{DBLP:conf/lpar/RegerS17}) 
two problems inherent to the solution: First, explicit axioms blow up the search space in the sense 
that a huge amount of consequences can additionally be generated. 
This happens since theory axioms are often repeatedly combined with certain clauses or among themselves, 
effectively creating cyclic patterns in the derivation.
Most of these consequences would immediately be classified as practically useless by humans. 
Second, many of the resulting consequences have small weight. 
This has the unfortunate effect that the age-weight clause selection heuristic \cite{otter3.0},
predominantly used by saturation-based theorem provers for guiding the exploration of the search-space,
often selects these theory-focused consequences.
This way the prover is getting lost in parts of the search space where the chance of finding a proof is low.


In this paper, we propose to limit the exploration of theory-focused consequences
by extending clause selection to take into account the amount of theory reasoning in the derivation of a clause.
Our solution consists of two parts.
First, we propose an efficiently computable feature of clauses, which we call \emph{th-distance}, that measures the amount of theory reasoning in the derivation of a clause (Sect.~\ref{sect:theory}).
Second, we turn to the general problem of incorporating a feature to a clause selection strategy.
There has been an ongoing interest in this problem \cite{DBLP:conf/cade/SchulzM16,DBLP:conf/cade/0001CV19,DBLP:conf/cade/Tammet19}.
We take inspiration from the layered clause selection approach presented in \cite{DBLP:conf/cade/Tammet19} 
and introduce the refined notion of \emph{multi-split queues}, which present a principled solution to the incorporation problem (Sect.~\ref{sect:layered}). 
We finally obtain a clause selection strategy for theory reasoning by instantiating multi-split queues with the feature \emph{th-distance}.
We implemented the resulting clause selection in the state-of-the-art saturation-based theorem prover~\vampire{} \cite{DBLP:conf/cav/KovacsV13},
and evaluate its benefits on a relevant subset of the \smtlib{} benchmark (Sect.~\ref{sect:exper}).
%
\paragraph{\bf Related work.}
There are different approaches to adding support for theory reasoning to saturation-based theorem provers,
either by extending the prover's inference system with dedicated inference rules
\cite{ordered-chaining-bg,coming-to-terms,DBLP:conf/cpp/KotelnikovKRV16,extensionality-resolution} 
or using even more fundamental design changes 
\cite{GCAI2016:AVATAR_Modulo_Theories,hierarchical-spass,hierachical-beagle,unification-with-abstraction-vampire}. 
While such solutions can result in very efficient reasoning procedures, their development is incredibly challenging and their implementation is a huge effort.
As a result, only a few theories are covered by such approaches, in contrast to our technique, which applies to arbitrary theories.
In particular, our technique can be used by non-experts on custom theory-domains coming from applications for which no dedicated solution exists.
Our work has similar motivation to \cite{DBLP:conf/lpar/RegerS17}, 
where the authors use the set-of-support strategy \cite{Wos:1965:ECS:321296.321302}
to limit the amount of reasoning performed with pure theory consequences.
However, unlike our technique, they do not impose any limit on clauses whose derivation contains at least one non-theory-axiom. 
\paragraph{\bf Contributions.} The summarized contributions of this paper are:
\begin{itemize}
   \item A new approach for building clause selection strategies from clause features, based on \emph{multi-split queues}.
   \item A new clause selection strategy for theory reasoning based on the instantiation of multi-split queues with the \emph{th-distance}-feature measuring the amount of theory reasoning in the derivation of a clause. Our solution applies to arbitrary theories and does not require fundamental changes to the implementation of clause selection.
   \item An implementation of the introduced clause selection strategy in the state-of-the-art theorem prover~\vampire{}.
   \item An experimental evaluation confirming the effectiveness of the technique, by improving on the existing heuristics 
   by up to \SI{37}{\percent} on a relevant set of benchmarks.
\end{itemize}

\section{Layered Clause Selection}

\label{sect:layered}


We assume the reader to be familiar with the saturation-based theorem proving technology (see, e.g.~\cite{Overbeek:1974:NCA:321812.321814,DBLP:books/el/RV01/BachmairG01}) 
and, in particular, with clause selection, the procedure for deciding, at each iteration of a saturation algorithm,
which of the currently passive clauses to next select for activation, i.e.~for participation in inferences with
the previously activated clauses. To agree on terminology, we start this section by recalling clause selection by age and weight.
We then move on to explaining layered clause selection.

The two most important features of a clause for clause selection are 
1) its \emph{age}, typically implemented using an ever-increasing ``date of birth'' timestamp,
and 2) \emph{weight}, which refers to the number of symbols occurring in the clause.
A theorem prover prefers to select clauses that are old, which implicitly corresponds
to a breadth-first search strategy, and clauses that are light,
which is a form of best-first search (clauses with few symbols are cheaper to process,
tend to be stronger simplifiers, and are intuitively closer to the ultimate target, the empty clause).
In practice, the best performance is achieved by combining these two criteria \cite{otter3.0,DBLP:conf/cade/SchulzM16}.
This is achieved by storing the passive clauses in two \emph{queues}, one sorted by age and the other by weight,
and setting a ratio to specify how the selection alternates between picking from these two queues.


\paragraph{\bf Layered Selection.}
In the system description of GKC \cite{DBLP:conf/cade/Tammet19}, 
Tammet describes an idea of using two layers of queues to organise clause selection.
The first layer relies on the just-described combination of selection by age and weight.
In the second layer, clauses are split into disjoint groups using a certain property
(e.g., ``being derived from the \emph{goal} or not'' could define two groups),
each group is represented by two sub-queues of the first layer,
and the decision from which group to select the next clause
is dictated by a new second-layer ratio. 
Although Tammet does not expand much on the insights behind using the layered approach,
he reports it highly beneficial for the performance of GKC.
In our understanding, the additional layer (in principle, there could be more than two) 
provides a clean way of incorporating into clause selection a new notion of what a preferred clause
should be, without a priori disturbing the already established and tuned primary approach,
such as selection by age and weight.\footnote{
A known alternative \cite{DBLP:conf/cade/SchulzM16} is to adapt the formula for computing weight to include a term
for penalising bad clauses 
and still rely on selection by age and this new refined notion of weight. 
(See also the \emph{non-goal weigth coefficient} in \cite{Vampire2019:Aiming_for_Goal_with}.)} 

Our preliminary experiments with the idea (instantiated with the derived-from-the-goal property)
found it useful, but not as powerful as other goal-directed heuristics 
in \vampire{}. In particular, finding a universally good ratio between the ``good'' clauses and the ``bad'' ones seemed hard.
What we propose here instead (and what also led in our experiment to a greater performance gain)
is to instead organise the clauses into groups with ``good'' ones and ``all''.
Here the second group contains all the passive clauses and essentially represents a fallback to the original single-layer strategy.
The advantage of this new take on layered selection is that a bad clause is only selected if
1) it is time to try a bad clause according to the second-layer ratio and 2) the best bad clause 
is also the current overall best according to the age-weight perspective.
This makes picking a good second-layer ratio much easier. 
In particular, one can ``smoothly'' move (by changing the second-layer ratio) from a high preference for the ``all'' second-layer queue towards selecting more ``good'' clauses 
without necessarily having to select any ``bad'' ones.

\paragraph{\bf Multi-split Queues.} We propose multi-split queues 
to realize layered selection with second layer groups defined by a real-valued clause feature.
\begin{definition}
Let $\mu$ 
be a real-valued clause evaluation feature such that preferable clauses have low value of $\mu(C)$.
Let the \emph{cutoffs} $c_1,\ldots,c_k$ be monotonically increasing real numbers with $c_k = \infty$,
and let the ratio $r_1:\ldots:r_k$ be a list of positive integer values.
These together determine a layered selection scheme with $k$ groups 
$\mathcal{C}_i = \{C | \mu(C) \leq c_i \}$ for $i = 1,\ldots,k$, such that we select
from the $i$-th group with a frequency $r_i / (\Sigma_{j=1}^k r_j)$.
\end{definition}
It is easy to see that multi-split queues generalise the binary ``good'' vs ``all'' arrangement,
since, thanks to monotonicity of the cutoffs, we have $\mathcal{C}_i \subseteq \mathcal{C}_{i+1}$.
Moreover, since $c_k=\infty$, $\mathcal{C}_k$ will contain all the passive clauses.

\section{Theory part} \label{sect:theory}


In this section, we instantiate the idea of multi-split queues from Sect.~\ref{sect:layered} 
with a concrete clause evaluation feature, which measures the amount of theory reasoning in the derivation of a clause.
We assume that the initial clauses given to the saturation algorithm, which we simply refer to as \emph{axioms},
consists of non-theory axioms obtained by clausifying the input problem
and theory axioms added to facilitate theory reasoning.

We start by defining the fraction of theory reasoning in the derivation of a general clause.
This relies on counting the number of theory axioms, resp.~the number of all axioms, in the derivation-tree using running sums.

\begin{definition}
   For a theory axiom $C$, define both $\thax(C)$ and $\allax(C)$ as $1$. For a non-theory axiom $C$, define $\thax(C)$ as $0$ and $\allax(C)$ as $1$. For a derived clause $C$ with parent clauses $C_1, \dots, C_n$, define $\thax(C)$ as $\sum_i \thax(C_i)$ and $\allax(C)$ as $\sum_i \allax(C_i)$.
   Finally, we set $\ratio(C) := \thax(C) / \allax(C)$.
\end{definition}

Assume now that for a given problem we expect (based on domain knowledge and experience) 
the fraction of theory reasoning in the final refutation $\ratio(\bot)$ to be at most $1 / d$, for a positive integer $d$.
Our clause evaluation feature \emph{th-distance} measures how much $\ratio(C)$ 
exceeds the expected ``maximally allowed'' fraction $1 / d$. 
More precisely, \emph{th-distance} counts the number of non-theory axioms which the derivation of $C$ would 
additionally need to contain to achieve a ratio $1 / d$.
\begin{definition}
   The \textit{th-distance}$\,:\mathit{Clauses}\to \mathbb{N}$ is defined as
   $$\text{\textit{th-distance}}(C) := \mathit{max}(\thax(C) \cdot d - \allax(C), 0).$$
\end{definition}

Our heuristic is based on the idea that a clause with small \emph{th-distance} is more likely to contribute to the refutation than a clause with high \emph{th-distance}. 
We therefore want to ensure that clause selection focuses on selecting clauses $C$ with a low value \emph{th-distance}$(C)$. 
We realize this with the multi-split queues (see Sect.~\ref{sect:layered}),
instantiating the clause evaluation feature $\mu$ by \emph{th-distance},
resulting in a second layer clause selection strategy with parameters $d$,\,
$c_1,\ldots,c_k$ and 
$r_1:\ldots:r_k$.

\section{Experiments} \label{sect:exper}

We implemented the heuristic described in Section~\ref{sect:theory} in 
\vampire{} (version 4.4).
Our newly added implementation consists of about 900 lines of C++ code and is compatible with 
both the LRS saturation algorithm \cite{DBLP:journals/jsc/RiazanovV03} and~\avatar{} \cite{DBLP:conf/cav/Voronkov14}.

For evaluation, we used the following subset of the most recent version (as of January 2020) of SMTLIB \cite{BarFT-SMTLIB}:
We took all the problems from the sub-logics that contain quantification and theories, such as LIA, LRA, NRA, ALIA, UFDT, \ldots\ 
except for those requiring bit-vector (BV) or floating-point (FP) reasoning, currently not supported by \vampire{}.
Subsequently, we excluded problems known to be satisfiable 
and those that were provable using \vampire{}'s default strategy in \SI{10}{\second} either without adding theory axioms or while performing clause selection by age only.
This way, we obtained \num{20795} problems.\footnote{A list of the selected problems 
along with other information needed to reproduce our experiments can be found at \url{https://git.io/JvqhP}.}



\begin{table}
\caption{Comparing clause selection strategies on \vampire{}'s default configuration.}
\label{tab:tuning}
\centering
\setlength{\tabcolsep}{5pt}
\begin{tabular}{ccccccc}
strategy & $d$-value & cutoffs & ratio & refuted & $\Delta$base & $\Delta$base\% \\
\hline
\texttt{default} & -- & -- & -- & \num{886} & 0 & 0.0 \\
\hline
\texttt{layered2} & 10 & $23, \infty$ & 33:8 & \num{1112} & 226 & 25.5 \\
\texttt{layered3} & 7 & $0, 30, \infty$ & 16:8:1 & \num{1170} & 284 & 32.1 \\
\texttt{layered4} & 8 & $16,41,59, \infty$ & 84:9:2:2& \num{1176} & 290 & 32.7
\end{tabular}
\end{table}



As a first experiment, we compared the number of problems solved in \SI{10}{\second} by the default strategy\footnote{The default strategy
uses \avatar{} \cite{DBLP:conf/cav/Voronkov14},
the LRS saturation algorithm \cite{DBLP:journals/jsc/RiazanovV03} and an age-weight ratio of 1:1.}
and its various extensions by multi-split queues defined in Sect.~\ref{sect:theory}.\footnote{The experiment was run on our local server with Intel Xeon \SI{2.3}{\giga\hertz} processors.}
The $d$-value, cutoffs and ratio values for the heuristic were selected by educated guessing and randomised hill-climbing.
Table~\ref{tab:tuning} lists results of the best obtained configurations.
It can be seen that already with two second layer queues a substantial improvement of \SI{25.5}{\percent} over the default
is achieved. Moreover, while it is increasingly more difficult to choose good values for the many parameters
defining a configuration with multiple queues, their use further significantly improves the number of problems solved.


\begin{table}
\caption{Comparing clause selection strategies on \vampire{}'s portfolio configuration.}
\label{tab:schedule}
\centering
\setlength{\tabcolsep}{5pt}
\begin{tabular}{ccccccc}
strategy & $d$-value & cutoffs & ratio & refuted & uniques \\
\hline
\texttt{SMTCOMP2019} & -- & -- & -- & \num{5479} & \num{194} \\
\texttt{SMTCOMP2019+layered4} & 8 & $16,41,59, \infty$ & 84:9:2:2&  \num{5629} & \num{344}
\end{tabular}
\end{table}



In a second experiment,\footnote{The second experiment was run on the StarExec cluster \cite{starexec} with \SI{2.4}{\giga\hertz} processors.} 
we ran \vampire{}'s strategy schedule for SMTCOMP 2019 \cite{SMTCOMP}
on our problems and also the same schedule additionally imposing the most successful second-layer clause selection scheme \texttt{layered4} from the first experiment.
The time limit was \SI{500}{\second} per problem. Table~\ref{tab:schedule} shows the results.

We can see that the version with second-layer queues improved over the standard schedule by 150 solved
problems. This is a very significant result, suggesting the achieved control 
of theory reasoning is incredibly helpful. 
Moreover, one should keep in mind that strategies in a schedule are carefully selected to complement each other
and even locally good changes in the strategies often destroy this complementarity (cf., e.g., \cite{DBLP:conf/cade/RawsonR19,DBLP:conf/lpar/RegerS17}).
In our case, however, we achieve an improvement despite this looming negative effect.
Finally, it is very likely that a new schedule, constructed while taking our new technique into account,
will be able to additionally cover some of the \num{194} problems currently only solved by the unaltered schedule.

\section{Conclusion}
We introduced a new clause selection heuristic for reasoning in the presence of explicit theory axioms. 
The heuristic is based on the combination of multi-split queues and a new clause-feature measuring the amount of theory reasoning in the derivation of a clause.
Our experiments show that the new heuristic significantly improves the existing state-of-the-art clause selection strategy.
As future work, we want to extend layered clause selection with new clause-features and 
combine it with the machine-learning-based approach in the style of ENIGMA \cite{DBLP:conf/cade/ChvalovskyJ0U19}.







\bibliographystyle{plain}
\bibliography{bib}

\end{document}